\documentclass[floats,floatfix,amssymb,prl,twocolumn,superscriptaddress,nofootinbib]{revtex4-1}
		
\usepackage{amssymb,amsmath,verbatim,mathtools,needspace,enumitem,etoolbox,graphicx,physics,microtype,afterpage,bm}
\usepackage[dvipsnames, usenames]{xcolor}
\definecolor{linkcolor}{rgb}{0.0,0.3,0.5}
\usepackage[unicode, colorlinks=true, linkcolor=linkcolor, citecolor=linkcolor, filecolor=linkcolor,urlcolor=linkcolor, pdfusetitle]{hyperref}
\usepackage[all]{hypcap}
\usepackage[T1]{fontenc}
\usepackage[utf8]{inputenc}
\usepackage{tabularx}
\usepackage{float}
\usepackage{lmodern}
\allowdisplaybreaks
\usepackage{tikz}
\usepackage{color}
\usepackage{framed}
\usepackage{hyperref}
\hypersetup{colorlinks, citecolor=bluscuro, linkcolor=black, urlcolor=bluscuro}
\definecolor{rossos}{cmyk}{0,1,1,0.55}
\definecolor{bluscuro}{rgb}{0.15, 0.2, .85}
\definecolor{bluchiaro}{cmyk}{1,.3,0.,0.1}
\definecolor{ForestGreen}{rgb}{0.13, 0.55, 0.13}

\newcommand{\be}{\begin{equation}}
\newcommand{\ee}{\end{equation}}

\newcommand{\tn}{\textnormal}
\def\lsim{\mathrel{\rlap{\lower4pt\hbox{\hskip0.5pt$\sim$}}
    \raise1pt\hbox{$<$}}}         %less than or approx. symbol
\def\gsim{\mathrel{\rlap{\lower4pt\hbox{\hskip0.5pt$\sim$}}
    \raise1pt\hbox{$>$}}}         %greater than or approx. symbol

\begin{document}

\title{Detecting Subsolar-Mass Primordial Black Holes \\in Extreme Mass-Ratio Inspirals with LISA and Einstein Telescope}

\author{Susanna Barsanti}
\email{susanna.barsanti@uniroma1.it}
\affiliation{Dipartimento di Fisica, Sapienza Università 
	di Roma, Piazzale Aldo Moro 5, 00185, Roma, Italy}
\affiliation{INFN, Sezione di Roma, Piazzale Aldo Moro 2, 00185, Roma, Italy}

\author{Valerio De Luca}
\email{valerio.deluca@unige.ch}
\affiliation{D\'epartement de Physique Th\'eorique and Centre for Astroparticle Physics (CAP), Universit\'e de Gen\`eve, 24 quai E. Ansermet, CH-1211 Geneva, Switzerland}
\affiliation{Dipartimento di Fisica, Sapienza Università 
di Roma, Piazzale Aldo Moro 5, 00185, Roma, Italy}

\author{Andrea Maselli}
\email{andrea.maselli@gssi.it}
\affiliation{Gran Sasso Science Institute (GSSI), I-67100 L'Aquila,  Italy}
\affiliation{INFN, Laboratori Nazionali del Gran Sasso, I-67100 Assergi, Italy}

\author{Paolo Pani}
\email{paolo.pani@uniroma1.it}
\affiliation{Dipartimento di Fisica, Sapienza Università 
	di Roma, Piazzale Aldo Moro 5, 00185, Roma, Italy}
\affiliation{INFN, Sezione di Roma, Piazzale Aldo Moro 2, 00185, Roma, Italy}

% \date{\today}

\begin{abstract}
Primordial black holes possibly formed in the early universe could provide a significant fraction of the dark matter and would be unique probes of inflation. 
A smoking gun for their discovery would be the detection of a subsolar mass compact object. We argue that extreme mass-ratio inspirals will be ideal to search for subsolar-mass black holes not only with LISA but also with third-generation ground-based detectors such as Cosmic Explorer and the Einstein Telescope.
These sources can provide unparalleled measurements of the mass of the secondary object at subpercent level for primordial black holes as light as ${\cal O}(0.01)M_\odot$ up to luminosity distances around hundred megaparsec and few gigaparsec for LISA and Einstein Telescope, respectively, in a complementary frequency range. This would allow claiming, with very high statistical confidence, the detection of a subsolar-mass black hole, which would also provide a novel (and currently undetectable) family of sources for third-generation detectors.
\end{abstract}

\maketitle

% \section{Introduction}
%----------------------------------------------------------------------------------------------------
\noindent{{\bf{\em Introduction.}}}
%----------------------------------------------------------------------------------------------------
Primordial black holes~(PBHs) were proposed more than 50 years ago~\cite{Zeldovich:1967lct,Hawking:1974rv,Chapline:1975ojl,Carr:1975qj} as BHs possibly formed across a vast mass range during the radiation-dominated era from the collapse of very large inhomogeneities~\cite{Ivanov:1994pa,GarciaBellido:1996qt,Ivanov:1997ia,Blinnikov:2016bxu}. 
Besides being unique messengers for inflation, in certain mass ranges PBHs could comprise the entirety of the dark matter, and could seed supermassive BHs at high redshift~\cite{Volonteri:2010,Clesse:2015wea, Serpico:2020ehh}.
For these reasons, a robust PBH detection would have dramatic consequences for astrophysics, cosmology, gravitation, and particle physics.

However, detecting PBHs has proved to be extremely challenging. On the one hand, various constraints exist for the fraction of PBHs in dark matter~\cite{Carr:2020gox}. On the other hand, in certain allowed mass ranges it is hard to disentangle the effect of a PBH from the astrophysical foreground. A notable example are the BH mergers detected by current and future GW interferometers.
In recent years a considerable effort has been put to understand whether (at least a fraction of) the GW events detected by LIGO-Virgo so far~\cite{LIGOScientific:2018mvr, LIGOScientific:2020ibl} are of primordial origin~\cite{Bird:2016dcv,Sasaki:2016jop,Eroshenko:2016hmn, Wang:2016ana, Ali-Haimoud:2017rtz, Chen:2018czv,Raidal:2018bbj, Hutsi:2019hlw, Vaskonen:2019jpv, Gow:2019pok,Wu:2020drm,DeLuca:2020qqa, Hall:2020daa,Wong:2020yig,Hutsi:2020sol,DeLuca:2021wjr,Deng:2021ezy,Kimura:2021sqz, Franciolini:2021tla} (see Refs.~\cite{Sasaki:2018dmp,Green:2020jor} for reviews).
While PBHs could explain the recently detected mass-gap events (GW190814~\cite{Clesse:2020ghq} and GW190521~\cite{DeLuca:2020sae}) and a subpopulation of PBHs is statistically preferred against certain astrophysical population models in the latest GW catalogue~\cite{Franciolini:2021tla}, confidently claiming that a BH merger is of primordial origin is much more challenging.
Attempts have been made for single-event detections using Bayesian model selection based on astrophysically or primordial-motivated different priors~\cite{Bhagwat:2020bzh}, whereas catalogue analyses could use the peculiar mass-spin-redshift distributions predicted for PBH binaries~\cite{Raidal:2018bbj,DeLuca:2020qqa} or perform population studies~\cite{Hall:2020daa,Wong:2020yig,Hutsi:2020sol,DeLuca:2021wjr,Franciolini:2021tla}. Unfortunately, none of these strategies seem able to give irrefutable evidence due to uncertainties in both PBH and astrophysical models~\cite{Franciolini:2021tla}.
Future third-generation~(3G) detectors such as Cosmic Explorer~(CE)~\cite{Reitze:2019iox} and Einstein Telescope~(ET)~\cite{Hild:2010id} could detect several PBH mergers at redshift $z>30$, where astrophysical-origin mergers should not occur~\cite{Koushiappas:2017kqm,DeLuca:2021wjr}. However, redshift measurements for those cosmological sources are typically inaccurate and prior dependent~\cite{Ng:2021sqn}.

In this complex scenario, a promising road to disentangle PBHs from astrophysical ones would be detecting a subsolar-mass compact object, since astrophysical BHs are expected to be born with a mass larger than the Chandrasekhar one~\cite{Miller:2020kmv, Phukon:2021cus, DeLuca:2021hde, Pujolas:2021yaw}
(see Refs.~\cite{LIGOScientific:2018glc,LIGOScientific:2019kan,Nitz:2020bdb,Wang:2021qsu, Nitz:2021mzz, Nitz:2021vqh} for constraints on subsolar objects from current GW data).

In this work we argue that subsolar mass BHs can be identified with unparalleled statistical confidence level if they perform extreme mass-ratio inspirals~(EMRIs) around a supermassive BH as those detectable by the future Laser Interferometer Space Antenna~(LISA)~\cite{LISA:2017pwj} (see~\cite{Guo:2017njn, Kuhnel:2018mlr} for related studies), and also as EMRIs around intermediate-mass BHs which would provide a novel, currently undetectable, GW source for 3G ground-based detectors.~\footnote{See in particular Ref.~\cite{Nitz:2020bdb} for 
bounds on the merger rate of binary systems with large mass ratios and a subsolar mass component from the LIGO/Virgo data.}

% \section{Setup}
%----------------------------------------------------------------------------------------------------
\noindent{{\bf{\em Setup.}}}
%----------------------------------------------------------------------------------------------------
% \noindent%
Due to their tiny mass ratio $q=\mu/M\ll1$, EMRI evolution 
can be modelled within BH perturbation theory~\cite{Teukolsky:1973ha,Barack:2018yvs}, 
by studying the quasi-adiabatic orbital motion of a 
point-particle with mass $\mu$ (the secondary) around 
a much heavier BH with mass $M$ (the primary). 
At variance with standard EMRI studies, we shall consider that the secondary is sufficiently lighter than a solar mass, $\mu< M_\odot$.
We consider the leading-order adiabatic evolution, 
focusing on quasicircular, equatorial orbits around a Kerr 
BH and neglecting the spin of the secondary. The latter 
choice is motivated both because the spin of subsolar-mass 
PBHs is expected to be negligible~\cite{Mirbabayi:2019uph,DeLuca:2019buf,DeLuca:2020bjf},
and because measurements of the other waveform parameters 
are not significantly affected by the secondary 
spin~\cite{Barack:2003fp,Huerta:2011kt,Huerta:2011zi,Piovano:2020zin, Piovano:2021iwv}, the latter entering at first post-adiabatic order.

We use the BH Perturbation Toolkit~\cite{BHPToolkit} to 
solve Teukolsky equation with arbitrary precision and 
compute the total energy flux $\dot E$ emitted by the 
binary.
The adiabatic evolution of the inspiral is then driven by 
the emitted flux according to the evolution equations for 
the binary radius and phase
\be
\label{Evo}
\frac{dr}{dt}=-\dot{E}\frac{dr}{dE_\tn{\rm orb}}\quad\ ,\quad \frac{d\Phi}{dt}= \frac{M^{1/2}}{r^{3/2}+ \chi M^{3/2}}\ ,
\ee
where we focused on prograde orbits only, and $E_\tn{\rm orb}$ 
denotes the  binary orbital energy for a particle around a 
Kerr BH with mass $M$ and dimensionless angular momentum $\chi$ 
\cite{Hughes:1999bq}.
The initial conditions $(r_0,\Phi_0)$ to integrate Eqs.~\eqref{Evo} 
are chosen such that the secondary reaches an orbit within 
a distance, $r_\tn{plunge}$, of $0.1M$ from the 
innermost stable circular 
orbit~(ISCO) in a given observation time $T$. 

We then compute the corresponding GW signal using the 
quadrupole approximation~\cite{Barack:2003fp, Huerta:2011kt}, 
keeping into account the detector pattern functions, which
can be expressed in terms of the source orientation 
$(\theta_\tn{s},\phi_\tn{s})$ and spin direction 
$(\theta_\tn{l},\phi_\tn{l})$ in a solar barycentric 
frame, see Ref.~\cite{Nair:2018bxj} for ET and 
Refs.~\cite{Apostolatos:1994mx, Cutler:1997ta} for LISA 
(henceforth we focus on ET; results with CE~\cite{Reitze:2019iox} would be qualitatively similar). 
We also keep into account the phase modulation induced 
by the orbital motion~\cite{Babak:2006uv}, and an effective 
description of both the LISA and ET triangle configuration as a 
network of two L-shaped detectors, with the second detector
rotated by $45^\circ$ with respect to the first one.

The GW signal in the time domain is completely determined
by the following set of parameters: $\vec{\theta}=(\ln M,\ln \mu,\chi,\ln d,\theta_\tn{s},\phi_\tn{s},\theta_\tn{l},\phi_\tn{l},r_0,\Phi_0)$, 
where $d$ is the luminosity distance of the source.
The corresponding signal-to-noise ratio (SNR) can be 
computed from the GW strain $h$ as ${\rm SNR}=\langle h\vert h\rangle^{1/2}$, 
where we have defined the usual inner product as $\langle h_1\vert 
h_2\rangle=4\Re\int_{f_\tn{min}}^{f_\tn{\rm max}}\frac{\tilde{h}_1(f)\tilde{h}^\star_2(f)}{S_n(f)}df$,
in terms of the detector spectral density $S_n$, taken from the 
analytical fit of Ref.~\cite{Robson:2018ifk} for LISA (also including the confusion noise from unresolved white-dwarf binaries) and from Ref.~\cite{Hild:2010id} 
for ET-D.
For both detectors the minimum frequency is set by requiring 
the binary to spend a time $T$ to span the frequency band up to 
$f_\tn{\rm max}$ when the secondary reaches $r_\tn{plunge}$. We assume that 
EMRIs are observed for $T=1\,{\rm yr}$ and $T=1\,{\rm hr}$ by 
LISA and by ET, respectively.

In the limit of large SNR, the posterior distribution of the source 
parameters $\vec{\theta}$ can be approximated by a multivariate 
Gaussian distribution centered around the true values $\vec{\hat{\theta}}$ 
of the waveform parameters, with covariance ${\bf \Sigma} = {\bf \Gamma}^{-1}$, 
where $\Gamma_{ij}=\langle \frac{\partial h}{\partial \theta_i}\vert\frac{\partial h}{\partial \theta_j}\rangle_{\theta=\hat{\theta}}$ is the Fisher 
information matrix.
The statistical error on the $i$-th parameter is then 
given by $\sigma_i=\Sigma^{1/2}_{ii}$. In the limit of 
large SNR, the errors (and the inverse SNR) scale linearly 
with the luminosity distance of the source.

Due to the long and computationally expensive waveforms generated numerically in the time domain, along with their derivatives, Fisher matrices for EMRIs are characterised by large condition numbers, resulting in the need of high-precision numerical 
methods to compute the statistical errors accurately~\cite{Gair:2012nm}. To this aim we have adopted the 
same setup discussed in \cite{Maselli:2021men}. 
In particular we compute the exact GW fluxes and the Fisher matrix with high-precision numerics, which 
guarantees a stable evaluation of the covariance matrix.

\begin{figure*}[t]
	\centering
	\includegraphics[width=0.48\textwidth]{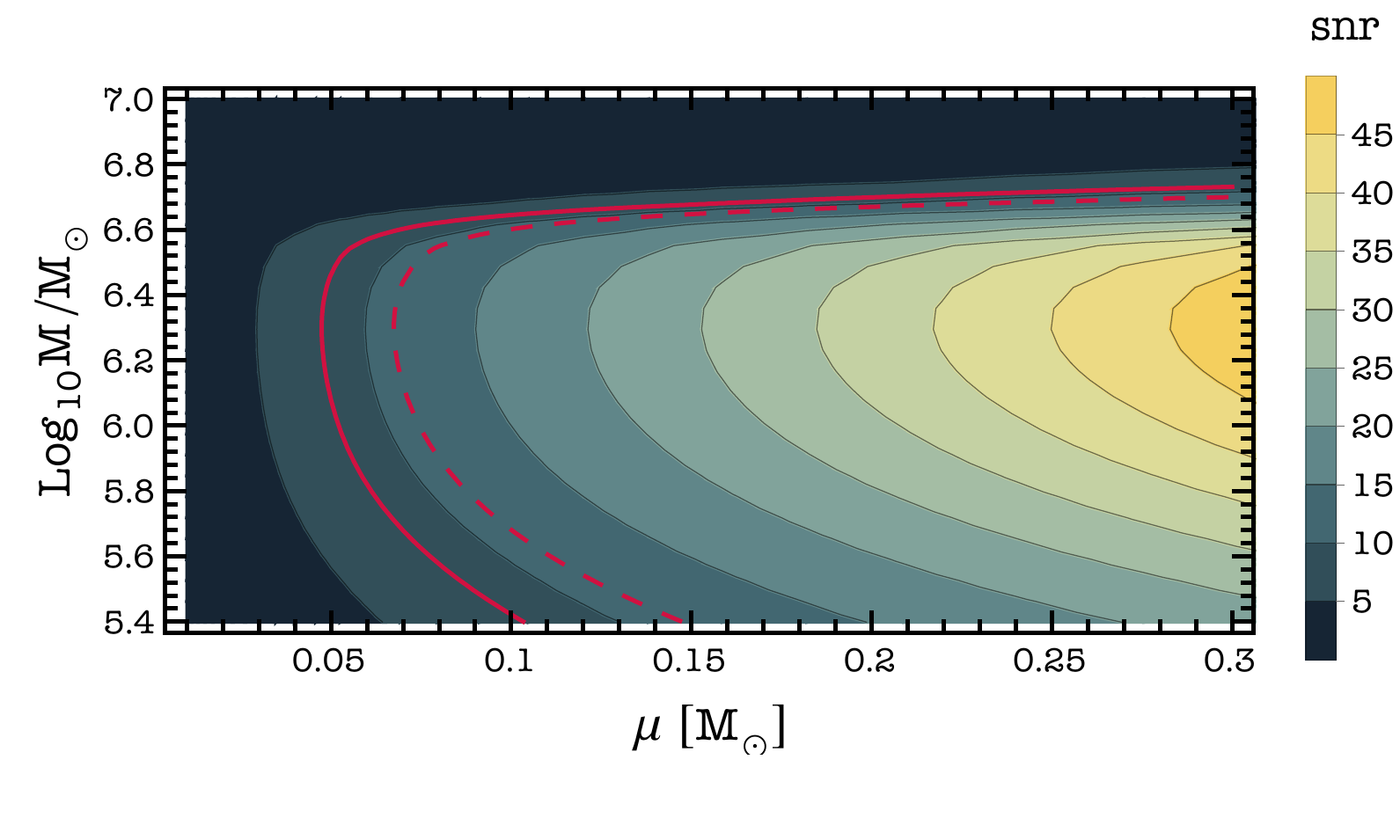}
	\includegraphics[width=0.48\textwidth]{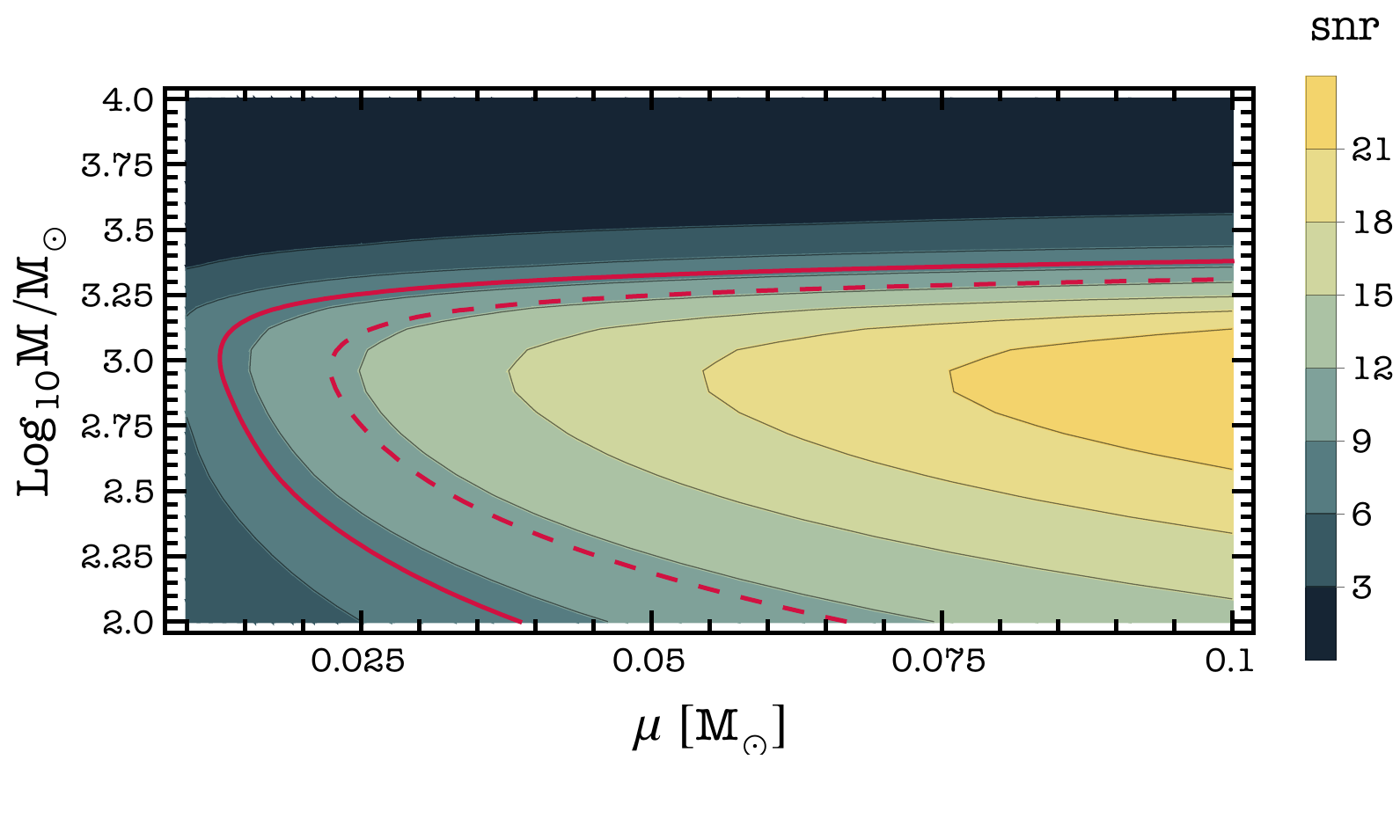}
	\caption{SNR for EMRIs with primary mass $M$ and subsolar secondary mass $\mu$ 
	observed by LISA (left) and ET (right). In both cases we have 
	fixed the primary spin to $\chi = 0.9$, assuming luminosity 
	distance $d = 100 \, {\rm Mpc}$ ($d = 1 \, {\rm Gpc}$) and signal 
	duration $T=1\,{\rm yr}$ ($T=1\,{\rm hr}$) for LISA (ET). The red solid (dashed) line identifies binaries with ${\rm SNR} = 8$ (${\rm SNR} = 11.3$).
	}
	\label{fig.SNR}
\end{figure*}

% \section{Results}
% \noindent
%----------------------------------------------------------------------------------------------------
\noindent{{\bf{\em Results.}}}
%----------------------------------------------------------------------------------------------------
We apply the numerical framework discussed above to investigate
the detectability of EMRIs with a subsolar-mass secondary. In the left panel of Fig.~\ref{fig.SNR} we 
show contour lines of fixed SNR for sources observed by LISA 
at $d=100$ Mpc, as a function of their component masses, assuming 
the spin of the primary $\chi = 0.9$\footnote{Hereafter we 
fix the source angles to $\theta_\tn{s}=\phi_\tn{s}=\pi/2$ 
and $\theta_\tn{l}=\phi_\tn{l}=\pi/4$, although we have also 
checked that random sampling their values does not significantly 
affect our numerical results.}. For a given $M$ the SNR 
decreases rapidly for smaller values $\mu$, since ${\rm SNR}\sim q^3$ 
in the EMRI limit.
Assuming a detection threshold ${\rm SNR}=8$ we find that, for the parameter space under consideration, 
binaries as distant as 
$\sim 500 \, {\rm Mpc}$ can be detected by LISA. For $d=100\,{\rm
Mpc}$ only secondary BHs heavier than $\sim 0.05 M_\odot$ 
can be potentially observed. However, EMRIs with $\mu\sim 0.1M_\odot$ 
feature ${\rm SNR}>8$ for a broad range of primary masses and can reach 
SNR as high as ${\rm SNR}\approx 40$ for $M\approx10^6M_\odot$ 
and $\mu\approx 0.3M_\odot$.
Assuming a larger threshold ${\rm SNR}= 11.3$, equal to the strength of a signal observed by two aligned detectors with ${\rm SNR}= 8$, one could potentially observe masses larger than $\sim 0.1 M_\odot$ and reach distances as large as $\sim 400 \, {\rm Mpc}$.

The corresponding analysis for ET is shown in the right panel 
of Fig.~\ref{fig.SNR}. Due to the different frequency band 
covered by ground-based interferometers, we focus here 
on intermediate-mass primaries with $M\in[10^2,10^4] M_\odot$, 
and $\chi=0.9$. 
Interestingly, ET's horizon for these sources is larger than LISA's horizon for their supermassive counterparts: in the right panel of Fig.~\ref{fig.SNR} we set $d=1\,{\rm Gpc}$. EMRIs in this peculiar mass range would
represent a new class of astrophysical sources for 3G interferometers, that can be observed with SNRs 
larger than those obtained for LISA binaries. At $d\sim100\,{\rm Mpc}$ ET 
would detect EMRIs with a secondary as small as 
$\mu\sim 10^{-2}M_{\odot}$ with ${\rm SNR}\gtrsim 20$ 
for $M\lesssim 2.5\times10^3M_\odot$. Our analysis also 
suggests that ET can observe subsolar mass BHs 
at cosmological distances: a typical system with $(M,\mu)=(10^3,10^{-1})M_\odot$
would be seen at the SNR threshold up to 
few gigaparsec.
This remarkable horizon is a peculiarity of the superior sensitivity (especially at low frequency) of 3G detectors such as ET relative to LIGO/Virgo. For the same sources shown in Fig.~\ref{fig.SNR}, the SNR in LIGO is smaller approximately by a factor $10$ to $100$ depending on the primary mass and spin, so all sources are well below the detectability threshold.

\begin{figure*}[t]
	\centering
	\includegraphics[width=0.49\textwidth]{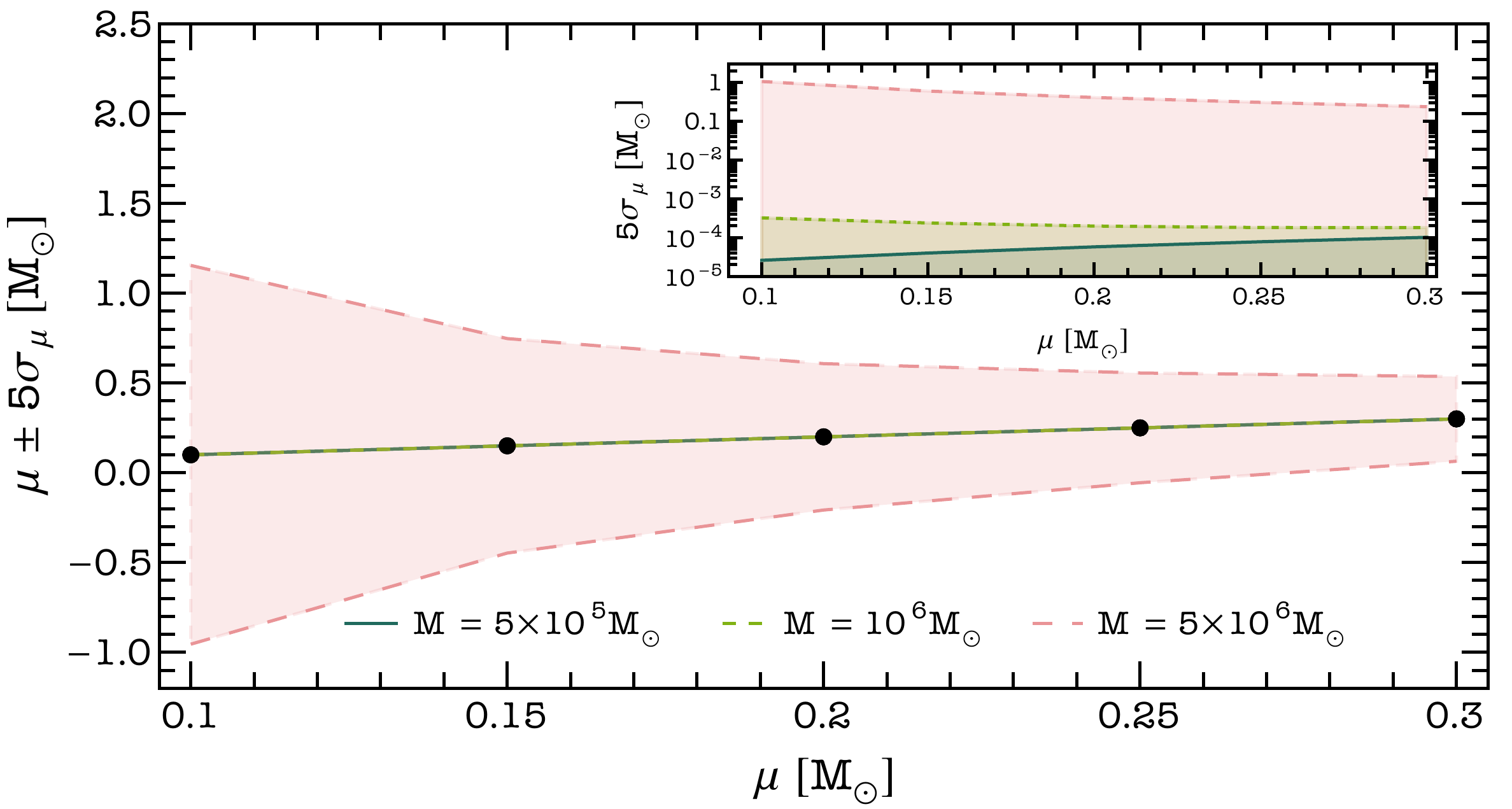}
	\includegraphics[width=0.49\textwidth]{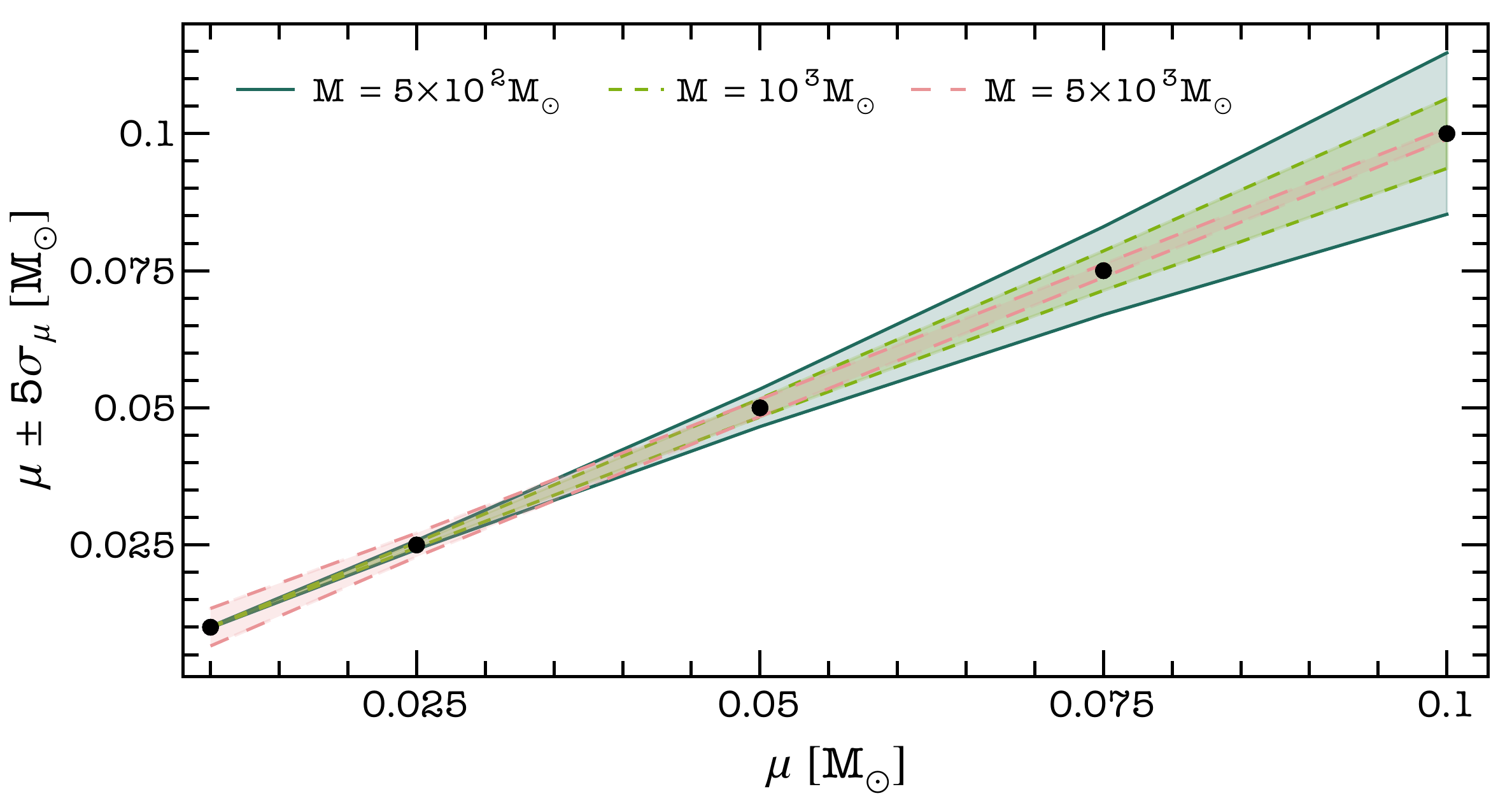}
	\caption{Injected values of the secondary mass (black dots) and their 5-$\sigma$ interval inferred for EMRIs observed by LISA (left) and ET 
	(right), assuming primary spin $\chi = 0.9$ and rescaling the 
	distance such that ${\rm SNR}=8$. Color bands identify binaries with different values of the primary mass. In the left panel two 
	of the bands are too narrow to be resolved, see inset for 
	the $5$-$\sigma$ (half) interval in logarithmic scale.}
	\label{fig.Fisher}
\end{figure*}

Despite their relatively low SNR, EMRIs are unique sources since their long orbital evolution provides measurements of the source parameters with unprecedented 
accuracy. This has been studied in details only for standard LISA EMRI sources with $\mu\gtrsim M_\odot$
~\cite{Barack:2003fp,Babak:2006uv,Babak:2006uv,Huerta:2011zi,Babak:2017tow,Katz:2021yft,Maselli:2021men,Piovano:2021iwv}.
Here we extend those analyses to subsolar secondaries and to ET for the first time. Projected constraints on the subsolar secondary mass are shown in Fig.~\ref{fig.Fisher} for LISA (left panel) and ET (right panel) 
and different binary configurations, assuming $\chi=0.9$ and a conservative value for the threshold ${\rm SNR}=8$ (larger thresholds will result in even smaller errors). 
Our results show that both ground and space interferometers are
able to measure the mass of a subsolar secondary component with subpercent precision in a large region of the detectable parameter space.
For LISA, all systems with $M\lesssim 10^6M_\odot$ would provide 
an indisputable identification of a subsolar component with 
$\mu\sim 0.1M_\odot$, with relative uncertainties well
below $0.1\%$. This would allow to exclude $\mu\gtrsim M_\odot$ for these systems at more than $5\sigma$ confidence level.

As shown in the right panel of 
Fig.~\ref{fig.Fisher}, this picture does not change qualitatively for the family of EMRIs that could be observed by ET. Overall, ET will allow to constrain values of the secondary BH mass 
down to $10^{-2}M_\odot$ for a wide range of primary mass ($M\in(10^2,10^4)M_\odot$), with 
relative errors $\sigma_\mu/\mu$ clustering below $10\%$ 
for $\mu \gtrsim 0.025M_\odot$, which would again allow excluding $\mu\gtrsim M_\odot$ at more than $5\sigma$ level.

The particular trend displayed by the errors in both 
panels of Fig.~\ref{fig.Fisher} is due to the dependence 
of the initial frequency $f_\tn{min}$ in terms of $\mu$. In particular, for large values of 
the primary mass, the initial frequency approaches the 
detector reach, while for smaller $M$, $f_\tn{min}$ grows 
as $\mu$ decreases for a given observing time $T$.

% \section{Discussion}
%----------------------------------------------------------------------------------------------------
\noindent{{\bf{\em Discussion.}}}
%----------------------------------------------------------------------------------------------------
% \noindent
We showed that, if subsolar BHs exist and form binaries 
around intermediate-mass and supermassive BHs, both LISA and 
ET are able to detect their inspiral up to distances of hundred 
megaparsec and few gigaparsec, respectively, and to identify a 
subsolar secondary mass in these sources at more than $5\sigma$ 
confidence level. 

An important question concerns whether a subsolar-mass detection, 
however robust, can be ascribed to some compact object other than 
a PBH.
White dwarfs and neutron stars are formed with masses respectively 
above $\approx 0.2M_\odot$~\cite{Kilic:2006as} and $\approx M_\odot$ 
in standard astrophysical scenarios. Furthermore, the Roche radius for 
a secondary white dwarf is larger than the ISCO of the primary when 
$\mu\gtrsim 0.002 \left(\frac{M}{10^3M_\odot}\right) M_\odot$, implying 
that white dwarfs and less compact stars (such as brown dwarfs) with 
$\mu\approx 0.2 M_\odot$ would be tidally disrupted before their 
plunge into a BH with $M\lesssim 10^5 M_\odot$. A similar conclusion holds also for brown dwarfs~\cite{Amaro-Seoane:2020zbo}, with characteristic masses $\approx 10^{-2} M_\odot$, which would be tidally disrupted before reaching the ISCO.
Therefore, a confident measurement of $\mu$ well below the solar-mass 
scale would necessarily imply new exotic physics.
Arguably, the most natural explanation for such a remarkable claim 
would be a population of subsolar PBHs which, given current microlensing 
constraints in that mass range~\cite{EROS-2:2006ryy}, could account 
for as much as a few percent of the dark matter~\cite{Carr:2020gox}.
In certain particle-dark-matter scenarios solar-mass BHs can form out 
of neutron star transmutation~\cite{Dasgupta:2020mqg, Giffin:2021kgb}, 
but lighter BHs can essentially be of primordial origin only (see 
however Ref.~\cite{Shandera:2018xkn} for models in which subsolar BHs 
are born out of dark sector interactions). 
Another possibility could be a subsolar exotic compact object~\cite{Cardoso:2019rvt}, for example a boson star~\cite{Guo:2019sns}, although also the latter should be compact enough not to be tidally disrupted.
In any case, detecting an EMRI with $\mu\ll M_\odot$ would imply new 
groundbreaking physics and should be included in the science case for 
fundamental physics with LISA~\cite{Barausse:2020rsu} and 3G GW 
detectors~\cite{Maggiore:2019uih}.

Given the impact that such a detection would have, our study should 
be extended in various directions. 
We have focused on a relatively conservative scenario in which the 
primary's spin is $\chi=0.9$. A faster-spinning primary would result in 
larger SNR, improving the detectability horizon by a factor of a few 
for $\chi=0.99$. Likewise, for $\chi=0.8$ the SNR decreases by a factor 
of a few relative to $\chi=0.9$, reducing the detectable parameter space. 
The errors on $\mu$ depend on $\chi$ less significantly.

Interestingly, not only would subsolar-mass BHs around intermediate 
ones provide a novel source for 3G detectors but --~given their mass 
ratio $q\sim 10^{-5}$ in the relevant parameter space~-- they would 
also allow exploiting the whole technology currently under development 
for standard EMRIs detectable by LISA, in particular first- and 
second-order self-force calculations~\cite{Pound:2012nt,Barack:2018yvs,Pound:2019lzj} 
and sophisticated parameter-estimation strategies to extract the EMRI signal from the whole LISA datastream~\cite{LISADataChallenge,Chua:2019wgs,Katz:2021yft}. Likewise, 
we have focused on circular equatorial orbits, but the estimated errors 
on the secondary mass for standard EMRIs are similar for eccentric and inclined 
orbits~\cite{Barack:2003fp,Huerta:2011kt,Babak:2017tow}, so we expect a comparable accuracy for EMRIs with $\mu\ll M_\odot$.
EMRI parameter estimation is a challenging and open 
problem~\cite{Huerta:2011kt,Babak:2017tow,Chua:2019wgs,Katz:2021yft}, 
which requires developing accurate waveform models, performing expensive 
statistical analysis, and also taking into account that the EMRI events 
in LISA might overlap with glitches, data gaps~\cite{Dey:2021dem}, and 
with several louder simultaneous signals from supermassive BH 
coalescences and other sources~\cite{Audley:2017drz,Chua:2019wgs,LISADataChallenge}. Detecting subsolar-mass EMRIs with LISA would face the same challenges.
In comparison, detecting primordial-origin EMRIs with 3G detectors 
should be less demanding, since the signal can be considerably shorter 
without compromising detectability and parameter estimation.
We have focused on a single 3G detector, but a network of ET plus 
one/two CE would further improve the overall SNR and the measurement 
errors.

The possibility of detecting subsolar BHs with 3G detectors gives further 
motivation to develop accurate waveform models for intermediate mass-ratio 
inspirals~(IMRIs), since for $\mu\sim 0.1 M_\odot$ and $M\sim (10^2-10^3) M_\odot$ 
the mass ratio $q\sim 10^{-4}-10^{-3}$ is in a range where finite-size 
effects are particularly relevant. In the IMRI regime a combination of 
numerical simulations, high-order post-Newtonian and self-force calculations, 
as well as effective one-body techniques, should be used for an accurate 
modelling and parameter estimation~\cite{Lousto:2010tb,Lousto:2010qx,Barack:2018yvs,Amaro-Seoane:2018gbb,Jani:2019ffg,vandeMeent:2020xgc}.

Another important question concerns the detection rates for 
subsolar BHs in EMRIs/IMRIs. Unfortunately, even the rates for ordinary 
EMRIs are rather uncertain~\cite{Babak:2017tow,Amaro-Seoane:2018gbb} 
and very little is known about the case of a subsolar-mass secondary.
By assuming that PBHs follow the dark-matter density distribution and 
by rescaling standard EMRI rates to subsolar masses, Ref.~\cite{Guo:2017njn} 
estimated that LISA could detect such sources as long as the fraction 
of PBHs in dark matter is a few percent. However, this conclusion relies 
on some approximations that should be carefully investigated.
The rates for subsolar BHs around intermediate-mass BHs are even more uncertain, since the population of intermediate-mass BHs is essentially unknown. However, the recent GW190521 event~\cite{LIGOScientific:2020iuh} shows that BHs with masses around and above hundred $M_\odot$ form at least as a result of previous mergers. From our analysis, a subsolar secondary with $\mu\sim{\rm few}\times 10^{-2} M_\odot$ around a primary with $M\sim {\rm few}\times 10^2 M_\odot$ could be detected and confidently identified by ET up to a few gigaparsec, and a two-detector ET+CE network would improve this horizon approximately by $60\%$ and would enlarge the detectable parameter space.

Finally, inspiralling binaries with subsolar components would emit a stochastic GW background which could be detected by LISA~\cite{Wang:2019kzb} and 3G detectors~\cite{Mukherjee:2021itf}. Resolving a primordial source would break the degeneracy between the PBH mass and abundance in the slope of the stochastic signal~\cite{Wang:2019kzb}, thus possibly unveiling the nature of the unresolved sources.

% \begin{acknowledgments}
%----------------------------------------------------------------------------------------------------
\noindent{{\bf{\em Acknowledgments.}}}
%----------------------------------------------------------------------------------------------------
We are very grateful to G. Franciolini and A. Riotto for insightful discussions and comments on the draft.
This work makes use of the Black Hole Perturbation Toolkit. 
Computations were performed at Sapienza University of Rome on the Vera cluster of the Amaldi Research Center.
V.DL. is supported by the Swiss National Science Foundation (SNSF), project {\sl The Non-Gaussian Universe and Cosmological Symmetries}, project number: 200020-178787. A.M. acknowledge support from the Amaldi Research Center funded by the MIUR program ``Dipartimento di Eccellenza" (CUP: B81I18001170001).
P.P. acknowledges financial support provided under the European Union's H2020 ERC, Starting 
Grant agreement no.~DarkGRA--757480, and under the MIUR PRIN and FARE programmes (GW-NEXT, CUP:~B84I20000100001), and support from the Amaldi Research Center funded by the MIUR program ``Dipartimento di Eccellenza" (CUP:~B81I18001170001).
% \end{acknowledgments}

\bibliography{draft}

\end{document}